\documentclass[useAMS,usenatbib]{mn2e}
\pdfoutput=1

\newcounter{MYtempeqncnt}
\usepackage{amssymb}
\usepackage{subfigure}
\usepackage[utf8]{inputenc}
\usepackage[english]{babel}
\usepackage{amsfonts}
\usepackage{amsbsy}
\usepackage{color}
\usepackage{multirow}
\usepackage{rotating}
\usepackage{slashbox}
\usepackage{ctable}
\newcommand{\be}{\begin{equation}}
\newcommand{\ee}{\end{equation}}
\newcommand{\bea}{\begin{eqnarray}}
\newcommand{\eea}{\end{eqnarray}}
\newcommand{\ben}{\begin{eqnarray}}
\newcommand{\een}{\end{eqnarray}}

\title[Constraints on (new) polynomial dark energy parametrizations: current results and forecasts]{SN and BAO constraints on (new) polynomial dark energy parametrizations: current results and forecasts}\author[Irene Sendra and Ruth Lazkoz]{Irene Sendra$^{1}$\thanks{E-mail:
irene.sendra@ehu.es} and Ruth Lazkoz$^{1}$\thanks{E-mail:ruth.lazkoz@ehu.es}\\
$^{1}$Fisika Teorikoa, Zientzia eta Teknologia Fakultatea, Euskal Herriko Unibertsitatea UPV/EHU, \\
644 Posta Kutxatila, 48080 Bilbao, Spain}

\begin{document}
\date{\today}
\pagerange{\pageref{firstpage}--\pageref{lastpage}} 

\maketitle

\label{firstpage}

\begin{abstract}
In this work we introduce two new polynomial parametrizations of dark energy and explore their correlation properties.  
The parameters to fit are the equation of state values at $z=0$ and $z=0.5$, which have naturally low correlation
and have already been shown to improve the popular Chevallier-Polarski-Linder (CPL) parametrization. 
We test our models with low redshift astronomical probes: type Ia supernovae and 
baryon acoustic oscillations (BAO), in the form of both current and synthetic data. Specifically, we present
simulations of measurements of the radial and transversal BAO scales similar to those expected in a BAO 
high precision spectroscopic redshift survey similar to  EUCLID. 
According to the Bayesian deviance information criterion (DIC), which penalizes large errors and correlations, we show that our models perform 
better than the CPL re-parametrization proposed by Wang (in terms of $z=0$ and $z=0.5$). This is due to 
the combination of a lower correlation and smaller relative errors.  The same holds for a frequentist perspective: 
our Figure-of-Merit is larger for our parametrizations. 
\end{abstract}


\section{Introduction}
More than a decade ago it was discovered that the universe is expanding with increasing velocity, and this 
fact was brought to light by type Ia Supernovae (SN) observations \citep{Perlmutter1999, Riess1998, Riess2004, Astier2006}. Nowadays the accelerated expansion of the universe stands as confirmed by several independent observations as the already mentioned SNeIa \citep{Hawkins2003, Goldstein2003, ReadHead2004, Tegmark2004, Spergel2007}; the measurements of cluster properties
as the mass, the correlation function and the evolution
with redshift of their abundance \citep{Eke1998, Viana2002, Bahcall2003, Bahcall2003a}; the optical
surveys of large scale structure \citep{Pope2004, Cole2005, Eisenstein2005}; the anisotropies in the
Cosmic Microwave Background (CMB) \citep{Bernardis2000, Spergel2003, Komatsu2011}; the cosmic
shear measured from weak lensing \citep{VanWaerbeke2001, Refregier2003} and
the Lyman$-\alpha$ forest absorption \citep{Croft1999, McDonald2005}.

This accelerated expansion can be explained postulating the existence of a ``new''
component in the universe, called ``dark energy" which would represent about between the $70\%-75\%$ the of the mass-energy of the universe and would counteract the effects of the gravitational attraction. The remaining $25-20\%$ would be composed by dark matter mainly and about an extra $5\%$ which would  account for the baryons, radiation and neutrinos. Although the existence of the dark energy is well established observationally, there is not a entirely compelling theoretical model which gives a theoretical physical framework within which the dark energy can be understood, and is of course compliant with all main observations.

Observational data agree rather well with the simplest model, $\Lambda$CDM, which puts down the acceleration to the presence of a cosmological constant $\Lambda$. This model has become the standard model of cosmology due to its simplicity and 
compliance with the data, but there are still some points that it can not account, e.g. the small value of the cosmological constant, which cannot be explained in terms of any of the known interactions. 

This situation leads to the proposal  of other settings which admit a slightly time-variable dark energy as the agent producing the the cosmic acceleration. Many 
such models have emerged along  theoretical avenues: quintessence \citep{Zlatev1999}, Chaplygin gas \citep{Bento2002}, modified gravity \citep{Nojiri2006}, 
holographic dark energy \citep{Li2004}, braneworld models \citep{Maartens2004}, $f(R)$ theories
\citep{Sotiriou2010}, theories with extra dimensions \citep{Lechtenfeld2002}, and quite a few others. 
Unfortunately none of them have emerged as definitive answers for the dark energy problem, so currently the situation is one of quick 
advances on the empirical front (regarding the quantity and quality of the data), whereas theory is somewhat mired in a jungle of alternatives. Since it 
is not fully clear what the observations should be compared with, a try and tested approach, borrowed from many areas of physics, is defining a 
parametrization of dark energy functions which, if well designed, allows to encapsulate all the observational information in a few numbers which can 
afterwards be compared with theoretical predictions. 

The most natural approach in this case is considering the dark energy equation of state. It is usually assumed that this quantity varies slowly 
with redshift and can be approximated by a fitting formula with a small number of free parameters. These parameters can be constrained comparing 
the ansatz with observations using an optimization procedure.  Choices are typically built upon intuition and prior information, but there is yet plenty 
of room for discussion and improvements. This is precisely the direction we follow in this work.

In section \ref{sec:param} we propose two new polynomial parametrizations of dark energy. We perform cosmological tests to compare them to the most 
popular dark energy parametrization in the literature, both in its classical fashion 
and a new, improved reformulation. We resort to several criteria to perform our comparison: in addition to 
the $\chi^2$ value, we also pay attention to the correlation of the coefficients, figure of merit, fractional errors on the free dark 
energy parameters and the deviance information criterion (DIC). The reason for an analysis from several perspectives is that we find $\chi^2$ to
 be an insufficiently informative tool, as it does not offer any reward upon some important improvements like tighter constraints or lower correlation.

 Our tests make use of Type Ia supernovae (SN) and baryon acoustic oscillations (BAO), which are so far the best representatives
in the categories of standard candles (objects with well determined intrinsic luminosity) and standard rulers (objects with 
well determined comoving size). Such probes provide us with distance measures related to the Hubble factor $H(z)$, and as they are low redshift  datasets, their suitability to constrain dark energy is strong, as this component of the cosmic soup has has began  to govern the evolution of the Universe just recently, according to most evidences. This combination has additional advantages:
SN measurements come in the form of luminosity distances and therefore give a smeared out information on $H(z)$ (two integrations are required) but still remain extremely useful because of the large number of measurements and their considerably good quality. On the other hand, BAO measurements, though currently scarce, involve $1/H(z)$ directly, so they are expected to favour sensitivity considerably, besides being of even better quality than SN data. In principle, one could also consider including CMB data, but typically they do not improve constraints on dark energy parameters significantly (WMAP7-year data alone constrain $w$ in quiessence models with about a $40\%$ error), and the dark matter density $\Omega_m$ is generally
the only parameter on which those data exert a strong impact. A good compromise between simplicity and 
advantages offered by CMB as regards $\Omega_m$ is the use of priors, and that is the approach we use.

Interestingly, we will not only use the latest observational data to obtain the constraints on the parameters of the models, but we will also consider mock data  simulating a forthcoming survey, as described in sections \ref{sec:data}. We present the  mock datasets for the two main measurable quantities expected from a line-of-sight, high-resolution spectroscopic baryon acoustic oscillations survey \citep{Per2010, euclid2,  Refregier2010, Beaulieu2010}, as it has done in a previous work \citep{EscamillaRivera:2011qb}.

We combine these data with synthetic pre-WFIRST (Wide-Field Infrared Survey Telescope) supernovae data to throw further light on the constraining power and suitability of the parametrizations proposed, so we are allowed to strengthen
our conclusions.

\section{Dark energy parametrizations}\label{sec:param}
The Friedman equations explain how an homogeneous and isotropic universe expands in the context of General Relativity (or generalizations). Considering such a universe is filled with several fluid components with pressures $p_i$ and energy
densities $\rho_i$, those equations will read
\ben
H^2 &\equiv&\left( \frac{\dot a}{a}\right)^2 =\frac{8\pi G}{3}\sum_i\rho_i-\frac{k}{a^2} \\
\frac{\ddot{a}}{a} &=& -\frac{4\pi G}{3}\sum_i(\rho_i+3p_i),
\een
where $a$ and $H$ are the scale and Hubble factor respectively.
At present, the two main such fluids are  dark matter and dark energy, the last being the  governor. 
Dark energy (de) enters our picture
in an effective way in the sense we do not appeal to any fundamental theory to deduce its behaviour and just let it be represented
by a phenomenological equation of state (EOS): 
\be
w(z)=\frac{\rho_{de}}{p_{de}}.
\ee
The importance  of the EOS is significant, because it determines the form  of the Hubble parameter $H(z)$ or any derivation of it which is necessary to obtain the observable quantities. Under the spatial flatness assumption ($k=0$)
\ben
\frac{H^2(z)}{H_0^2} &=&  
\Omega_m(1+z)^3+\Omega_{de}X(z)
\een
with 
\be
X(z)=\frac{\rho_{de}(z)}{\rho_{de}(0)}=\exp\left(3 \int_0^z\frac{1+w(z)}{1+z}dz\right),\label{xde}
\ee
and $\Omega_{de}=1-\Omega_m$.

Considering the huge number of contributions to the topic, and the knowledge so far gathered
it might seem hard to make improvements, but we believe some avenues 
opened by Wang in \citep{Wang2008a} are worth exploring and allow for little but valuable advances
towards parametrizations that make the best out of the data available. Questions like
the most convenient choice of the two dark energy parameters or the most suitable model comparison
criterion are worth being looked at once and again, particularly to build a more solid background
to make the most out of future data and their expected far better statistical value.

\subsection{Chevallier-Polarski-Linder parametrization}
The Chevallier-Polarski-Linder (CPL) parametrization, first discussed in \citep{Chevallier2001} and reintroduced in \citep{Linder2003}, defines the dark energy equation of state as
\be
w(z)=w_0+w_a\frac{z}{1+z},
\ee\label{CPLe}
where $w_0$ is the value of the dark energy equation of state today (i.e. at
redshift $z=0$).
This parametrization has been widely used because of its simplicity, sensitivity to observational data and because it is well behaved and bounded at high redshifts. Its adoption by the Dark Energy Task Force \citep{Albrecht2009} as a preferred parametrization has contributed
to its popularity. 

However, this parametrization has  shortcomings: its  flexibility to
determine cosmological parameters for some dark energy models with  rapid evolution
has been put in doubt recently, and on the other hand the second parameter is typically very poorly constrained thus losing
power of conviction about the conclusions to be drawn from it.

\subsection{Wang parametrization}
The quest to delineate the expansion history of Universe is expected to make big ambitious moves in the
future. As a result it is expected that  many more and better  observational data will be available. Obviously it is worth getting prepared to
making the best profit out of 
this avalanche of data to come, and this can be done by learning as much as possible from the data we already. 
Intuitively, in order to be able to discern which dark energy model is closest
to reality and to offer a sensible 
interpretation of the results
it seems necessary to consider dark energy parameters which offer clear advantages.
Apart from  having a clear physical meaning, for a  parameter to be eligible there should be previous hints or theoretical
grounds forecasting reasonably tight constrains, and as it is well know,  the parameter should be
the least possibly correlated to others. Suitability criteria for parameters along those grounds
can be sketched with the help of comparison tools rewarding for those nice features. The (frequentist) FoM favours low correlation whereas the (Bayesian) does not only favour that feature, but it also rewards for 
tight constraints also, so both these criteria (FoM and DIC) do poorly when high correlation is present in the dark energy 
parametrization one considers. In this respect, one must make a fair use of those statistical tools. For instance, if one
takes performs a  reparametrization of a certain scenario, the FoM and DIC as calculated for both cases will typically be 
direct indicators of improvements (or worsenings) in terms of correlation, but changes in the value of those quantities should not
be held in the same grounds as the changes occurring when  constraining the same single parametrization with two 
different datasets.

The CPL parametrization suffers from quite a significant correlation, besides the fact that constraints on one
of its parameters are typically large in percentual terms. 
 But on the other hand it has some nice features, so 
a convenient redefinition as concerns those two issues was put forward to yield an improved situation, although
the encoded information remains exactly the same. As mentioned, this was done in 
\citep{Wang2008a}, where a new dark energy description was given in
terms of its value at present, $w_0$, and at redshift $z=0.5$, $w_{0.5}$, was given.
Explicitly 
\be
w(z)=3w_{0.5}-2w_0+\frac{3(w_0-w_{0.5})}{1+z},\label{Wange}
\ee
was proposed, and this can be seen just as a rearrangement of the classic CPL parametrization.
This reformulation minorates the correlation between the parameters and 
allows to obtain tighter constrains along with a more transparent interpretation of the parameter estimation results, just
in the spirit of making an optimal use of observational data from future surveys.

The results in our paper reinforce the view that Eq. \ref{Wange} is a preferred way of exploiting the CPL parametrization, 
and on the other hand we suggest, with the help of two new parametrizations,  that the ($w_0,w_{0.5}$) couple is indeed a very good choice despite the specifics of the 
dark energy evolution, provided it is smooth enough and with bounded early asymptotic behaviour.

\subsection{Polynomial parametrizations}
\label{sec:conventional} 
The two proposals we are making are somewhat inspired on the one hand by the CPL 
parametrization  and on the other hand by an a proposal consisting in a expansion in powers of the quantity $(1+z)$ which emerged
naturally from the relationship between the redshift and the scale factor and showed computationally convenience. This 
second inspiring setup we are referring to was proposed for the first time in \citep{Weller:2001gf} in  the form
\be
w(z)=-1 + c_1 (1 + z) + c_2 (1 + z)^2.
\ee
However, this sort of parametrization poses problems at high redshifts, as $\vert w(z)\vert$ grows unboundedly with 
$z$ and one will either end up with a superphantom model or a superluminal one. This motivates considering generalizations
which are devoid of this pathology and may match or event surpass the nicety of CPL or its reformulation (recall Eq. \ref{Wange}).

Two possible routes that retain some similarity with the previous case but also some improvements arise from
\be
w(z)=-1 + c_1 (1 + f(z)) + c_2 (1 + f(z))^2.
\ee
with $f(z)$ a smooth and at the same time simple function
or the slightly more general
\be
w(z)=-1 + c_1 \,g_1(1 + f(z)) + c_2\, g_2(1 + f(z)),
\ee
where $g_1$ and $g_2$ are some smooth and simple functions as well.

\subsubsection{Conventional polynomial}
In the spirit of the first scheme above we propose
\be
f(z)=\frac{z}{1+z}
\ee
so we avoid high-redshift unboundedness by the same via as in CPL. Therefore
we have 
\be
w(z)=-1 + c_1 \left(1 + \frac{z}{1+z}\right) + c_2 \left(1 + \frac{z}{1+z}\right)^2,
\ee
which we dub conventional polynomial parametrization. If more compactness is desired
one can also write the latter as
\be
w(z)=-1 + c_1 \left(\frac{1+2z}{1+z}\right) + c_2 \left(\frac{1+2z}{1+z}\right)^2.
\label{conv}
\ee
It is convenient to leave out the constant term of value $-1$ from the computational point of view
because as one does not expect a large departure from a $\Lambda$CDM setting, it is reasonable to 
confine (at least initially) the parameter search region to $\vert c_1\vert <1$ and to $\vert c_2\vert <1$.

However, as we have discussed already it is desirable to fit parameters which are more or less
physically transparent 
and it in addition are just lightly correlated. At a first
stage we wish to compare our conventional polynomial parametrization with Wang's, so the best way to do so is to consider
exactly the same two dark energy parameters. Then, at a later stage, we will check whether the good
behaviour as correlation is concerned is shared by our parametrization. With those arguments in mind we 
reformulate the proposal made in Eq. \ref{conv} by letting
\bea
c_1=\frac{1}{4} (16 w_0-9
   w_{0.5}+7),
\\c_2= -3 w_0+\frac{9
   w_{0.5}-3}{4}.
\eea
This way, we move on to a scenario in which $w_0$ and $w_{0.5}$ are the parameters subject to estimation, 
which can be explicitly reconstructed using Eq. (\ref{xde}):
\bea &&X(z)=(1+z)^{\frac{3}{2} (-8 w_0+9 w_{0.5}+1)}\nonumber \\
&&e^{
   \left(\frac{3 z (w_0 (52 z+40)-9 w_{0.5} (5 z+4)+7
   z+4)}{8 (1+z)^2}\right)},~~
   \eea
   which consistently comes down to the $\Lambda$CDM case for $w_0=w_{0.5}=-1$.

For completeness and purposes related to model selection, it is convenient to compute and effective $w_a$ parameter for this
model. By analogy with the CPL case we define
\be
w_a=\lim_{z\to\infty}w(z)-w_0,
\ee
so in this case
\be
w_a=-5 w_0+\frac{9 }{2}w_{0.5}-\frac{1}{2}
\ee
is the result we get.
\subsubsection{Chebychev polynomial parametrization}
Now we want to make a further generalisation in the fashion of 
our general proposal above, by considering a bit more involved functions.
In this case we make use of Chebychev polynomials
(of the first kind), which have a significant role in most areas
of numerical analysis, as well as in other areas of Mathematics
(polynomial approximation, numerical
integration, and pseudo spectral methods for partial differential
equation, etc.)

Once again we have to take into account that we can not constrain accurately more than 
two dark energy parameters \citep{Linder2005}, thus we have cut the expansion at the second order.
Specifically we propose
\be
w(z)=-1+c_1T_1(1+f(z))+c_2T_2(1+f(z)),
\ee
with $T_n$ being the first kind Chebyshev polynomial of degree $n$
and $f(z)=z/(1+z)$ as before. 
A convenient presentation of that parametrization is 
\be
w(z)=-1 + c_1 \left(\frac{1+2z}{1+z}\right) + c_2 \left[2\left(\frac{1+2z}{1+z}\right)^2-1\right].
\ee
In this case too we switch to more amenable parameters and then let
\bea
c_1&=&\frac{1}{11} (23 w_0-9
   w_{0.5}+14)\\
c_2&=& -\frac{3}{11} (4 w_0-3
   w_{0.5}+1).
\eea
Then, one should simply resort to Eq. (\ref{xde}) to produce the whole scenario:
\bea
&&X(z)=(1+z)^{-\frac{3}{4} (52 w_0-45 w_{0.5}+7)}  \nonumber \\
&&e^{
   \left(\frac{3 z (w_0 (68 z+56)-9 w_{0.5}(6 z+5)+14
   z+11)}{4 (1+z)^2}\right)}.~~
\eea

For this first attempt
at depicting a dark energy dominated universe with Chebyshev polynomials in terms of those functions we
choose $w_0$ and $w_{0.5}$ as our parameters, but it it would be not surprising
than one could do better if some extra work was done in the direction of making
parameter correlation smaller. For the time being we just pursue to compare directly our polynomial proposal with a preferred
presentation of the CPL parametrization on the one hand and with our conventional polynomial proposal on the other hand.

In that direction and as it has been done before, we compute an effective $w_a$ parameter which in this case takes the form
\be
w_a=\frac{1}{11} (-49 w_0+45 w_{0.5}-4).
\ee

\subsubsection{Some additional remarks}
In the first place, these two new routes can be viewed as perturbations of $\Lambda$CDM, particularly, at low redshifts, which is the region  
most accurately described by the current data. If one wants to perturb $\Lambda$CDM in either the CPL or the Wang scenarios $w_0$ has 
to be anchored at $-1$ and then there is only one free parameter to play with. In contrast, our two new models may model two-parametric
departures from $\Lambda$CDM, and thus have more flexibility in principle.

Note that it would be possible to consider the two new parametrizations along with CPL(Wang)
if one let the parameter space have one more dimension. Indeed, if we let our parametrization be
of the form
\be
w(z)=b_1 + b_2 \left(\frac{1+2z}{1+z}\right) + b_3 \left(\frac{1+2z}{1+z}\right)^2,
\ee
then our conventional polynomial case would be obtained for $b_1=-1$, $b_2=c1$, and 
$b_3=c_2$; the Chebyshev case would follow from the choice $b_2=c_1$, $b_3=2c_2$ and $b_1=-(1+c_2)$; and
finally the CPL case would be obtained from $b_1=w_0-w_a$, $b_2=w_a$ and $b_3=0$.
\section{Observational data}\label{sec:data}
As dark energy is expected to have started to dominate at recent times, low redshift datasets are the obvious choice
to put the tighter constraints on each dynamics, whereas high redshift ones may be viewed as complementary. Thus
the combination of SN and BAO datasets, given their quality in both cases, and 
the considerable number of data points in the case of the SN, is an excellent choice given the state of the art.
Besides, new avenues on BAO \citep{Bassett2010} are to be open soon which will allow to exploit the tremendous
potential of this new astronomical tool towards constraining the main evolutionary features of dark energy. 
As we have already mentioned, one of our objectives is to introduce  new promising parametrizations as alternatives
to one of the commonest, but one of the other objectives is to forecast how the old parametrizations and our challengers
will cope with new data. 

The literature provides a large  number of papers where simulated supernovae data are used in the way we have just
mentioned, but to our knowledge synthetic baryon acoustic oscillations data have only been presented and
exploited in \citep{EscamillaRivera:2011qb}. This builds on considerable theoretical efforts in different forecast aspects
\citep{Blake2006}, which have crystallized in the package Initiative for Cosmology (iCosmo) (see \citep{Refregier2011}) and its 
BAO modules, which have allowed us to produce these mock data. We have modified and extended this  general purpose software 
to produce mock data from a line of sight, high-precision BAO spectroscopic survey  
as the one described in \citep{Per2010,euclid2} and pre-WFIRST supernovae data.

\subsection{Baryon Acoustic Oscillations}

Baryon Acoustic Oscillations (BAO) have emerged as a promising standard ruler in cosmology,
enabling precise measurements of the dark energy parameters with a
minimum of systematic errors \citep{Wang:2010gq,Blake2006}. These oscillations were originated before recombination
due to the density fluctuations created by acoustic waves generated by primordial perturbations. After recombination, photons decoupled and 
propagated freely leaving a signature of the primordial perturbations in the CMB temperature distribution. A similar but attenuated feature 
appears in the clustering of matter; the peaks and troughs of the acoustic waves gave rise to overdense regions of baryonic matter 
which imprint a correlation between matter densities at the scale of the sound horizon at recombination
\be
r_s(z_r)=\int_{z_{r}}^{\infty}\frac{c_s(z)}{H(z)}dz,
\ee
where $c_s$ is the sound speed \citep{Bassett2010, Percival2006}.
Cosmological Microwave Background (CMB) anisotropies provide an absolute physical scale for these baryonic peaks, but  these features 
can be also determined in a galaxy survey. By comparison of the absolute  value given by CMB and the observed location of the peaks of 
the two-point correlation function of the matter distribution, $\xi(z)$  given by a galaxy survey, one can obtain measurements of cosmological 
distance scales. Measuring the BAO scale from galaxy clustering in the transverse and radial directions yields measurements of $r(z)/r_s(z_{\rm{r}})$ and of $r_s(z_{\rm{r}})H(z)$
\citep{Blake2003,Hu2003,Seo2003}, where
\be
r(z)=\int_0^z{\frac{cdz'}{H(z')}}
\ee
is the comoving distance at redshift $z$.
\subsubsection{Percival et al.
}
In \citep{Percival2010} Gaussian values on the distance ratio,
$r_{s}(z_{\rm{drag}})/D_V(z)$, at redshifts $z=0.2$ and $z=0.35$, are given from the
measures obtained by combining the spectroscopic Sloan Digital Sky
Survey (SDSS) and the Two-Degree Field Galaxy Redshift Survey (2DFGRS)
data. This distance ratio represents the comoving sound horizon at the
baryon dragging epoch, $z_{\rm{drag}}$,
\begin{equation}
r_{s}(z_{\rm{drag}}) = c \int_{z_{\rm{drag}}}^{\infty} \frac{c_{s}(z)}{H(z)} \mathrm{d}z\; ,
\end{equation}
over the effective distance $D_{V}(z)$, defined as
\citep{Eisenstein2005} as 
\begin{equation}
D_{V}(z) = \left[ (1+z)^2 D_{A}^2 (z)\frac{c \, z}{H(z)} \right]^{1/3},
\end{equation}
$D_{A}$ being the angular diameter distance which takes the form
\begin{equation}
D_{A}(z) = \frac{c}{1+z} \int_{0}^{z} \frac{\mathrm{d}z'}{H(z')},
\end{equation}
in a flat universe containing only matter and dark energy.

In order to estimate the dark energy parameters in the context of Bayesian statistics, as detailed in Appendix \ref{ap:A}, we need a definition of the
$\chi^2$ which reflects the difference between the observational data and the values given by our models. In our case this requires giving an expression
for the comoving sound horizon at the baryon dragging epoch, and  we have used
the fitting formula proposed in \citep{Eisenstein1997}:
\be
r_{s}(z_{\rm{drag}})=153.5 \left(\frac{\Omega_b h^2}{0.02273}\right)^{-0.134}\left(\frac{ \Omega_m h^2}{0.1326}\right)^{-0.255}.
\ee

Now, taking into account the Gaussian values at $z=0.2$ and $0.35$
from the BAO data in \citep{Percival2010}, we can calculate
\be
\chi^2_{\rm BAO}=(v_i-v^{\rm BAO}_i)(\matrix{C}^{-1})^{\rm BAO}_{ij}(v_j-v^{\rm BAO}_j)
\ee
where
\bea
\textbf{v}=\left\{\frac{r_s(z_{\rm{drag}},\Omega_m,\Omega_b; \boldsymbol{\theta})}{D_V(0.2,\Omega_m; \boldsymbol{\theta})},\frac{r_s(z_{\rm{drag}},\Omega_m,\Omega_b; \boldsymbol{\theta})}{D_V(0.35,\Omega_m; \boldsymbol{\theta})}\right\},~~
\eea
\bea\textbf{v}^{\rm BAO}=\left(0.1905,0.1097\right)\eea and 
\begin{eqnarray}
\mathbf{C}^{-1}=\left(\begin{array}{cc}
30124  &  -17227\\
-17227  &  86977 \\
\end{array} \right),
\end{eqnarray}
being the inverse of the covariance matrix.

\subsubsection{High precision spectroscopic redshift surveys}

Future BAO surveys are expected to represent a real breakthrough in our 
knowledge of the dark energy. In 
\citep{Per2010, euclid2} is proposed a survey 
to measure spectroscopic redshifts for $6.1\times10^7$ luminous galaxies and clusters of galaxies out to redshift $z=2$ over 20000 $deg^2$, reaching a $dz/(1+z) < 0.001$, 
enough to resolve the BAO feature along the line of sight, and achieving much better dark energy constraints than their 
predecessors. 

\begin{figure*}

\includegraphics[width=0.4\textwidth]{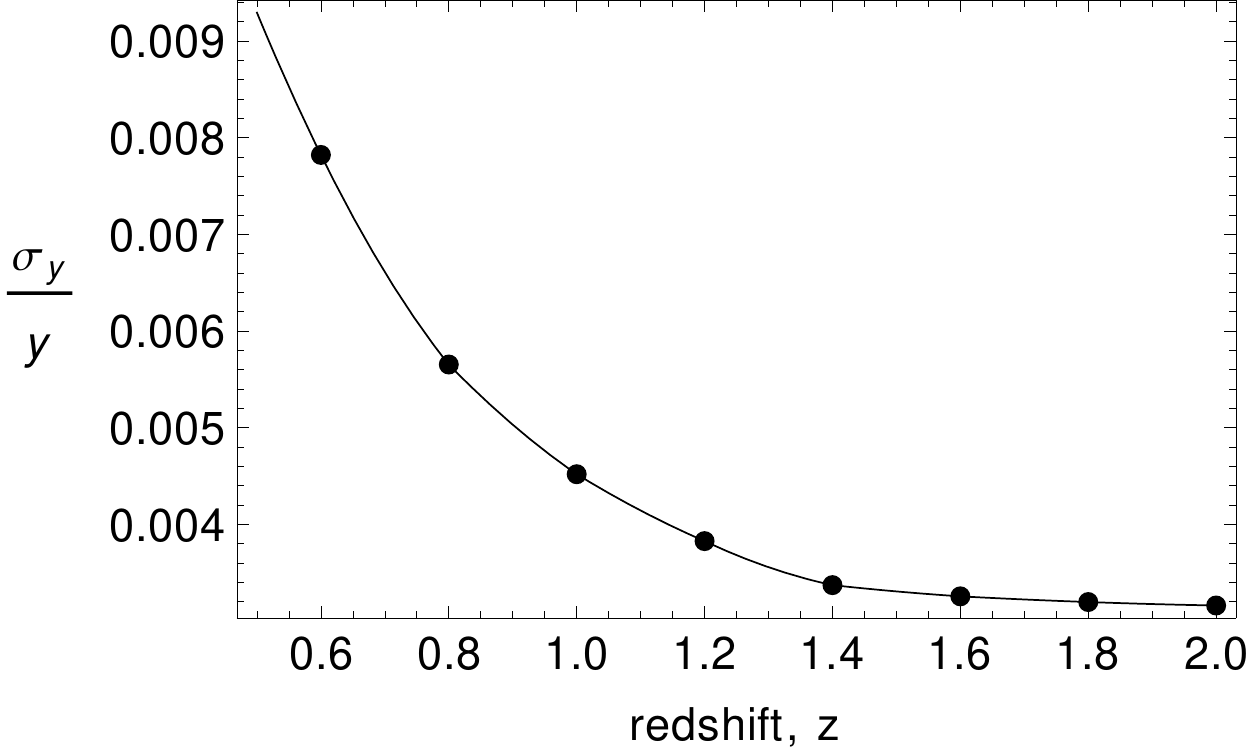}
\label{fig:yprima}
\includegraphics[width=0.4\textwidth]{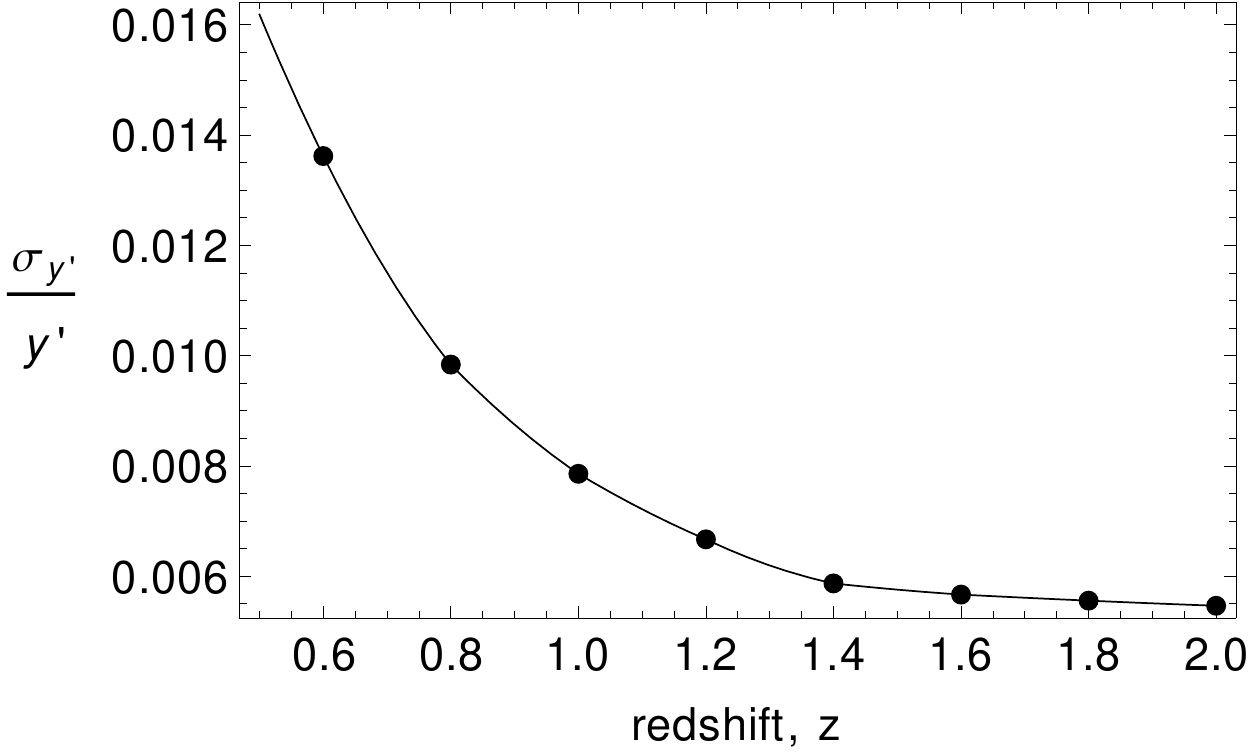}
\label{fig:yprimadata}
\caption{\label{fig:baodata} Fractional errors on synthetic BAO data for $y$ (a) and $y'$ (b).}
\end{figure*}

We have used the Initiative for Cosmology (iCosmo) software package to generate the BAO mock data. 
In this case are concerned with the 
following two signatures of the BAO peak:
\setlength\arraycolsep{0.1em}
\ben
y(z)&=&\frac{r(z)}{r_s(z_{\rm{r}})}, \\
y'(z)&=&\frac{r'(z)}{r_s(z_{\rm{r}})}=\frac{c/H(z)}{r_s(z_{\rm{r}})}.
\een
The publicly available code has built-in routines based on the universal BAO fitting
formulae for the diagonal errors on $y$ and $y'$ presented in
\citep{Blake2006}.
 Once we have made the proper modifications of the code to replace the defaults (Peacock and Dodds \citep{Peacock:1996ci} power
 spectrum and Smail et al. \citep{Smail:1994} galaxy distribution) with 
 the survey properties (those quoted above and besides $z_{med}\sim0.46$, redshift range $0.1<z<0.9$), we have 
 written extra codes to generate 
 a large number of normal random realizations around a fiducial model  which is specifically the  wcdm+sz+lens case from WMAP7-year
\citep{Komatsu2011}, which has $\Omega_m= 0.26\pm0.099$ and is phantom-like with $w= -1.12\pm0.43$.,
 Then after some reduction the synthetic BAO data presented in table
 \ref{tab:JPASdata} have been obtained. In addition, in the  the corresponding $\chi^2$ we have introduced priors based on the values of the matter
 and baryon density presented in \citep{Burigana2010} as a forecast analysis of Planck: using the result
 $\Omega_mh^2=0.1308\pm0.0008$ we construct a weak Gaussian prior, whereas
 with $\Omega_bh^2=0.0223$ we construct a fixed prior. In both cases we use $h=0.742$ as is given by \citep{Riess2009}. See as well
 Fig. \ref{fig:baodata} for a graphical account of the features
 of our BAO simulated data.

In this case we need an expression for the sound horizon at
recombination to constrain the dark energy parameters:
\begin{equation}
r_{s}(a_{r}) = \frac{c}{\sqrt{3}H_{0} \Omega_{m}^{1/2}} \int_{0}^{a_{r}} \frac{\mathrm{d}a}{(a+a_{eq})^{1/2}}
\frac{1}{(1+R)^{1/2}} \; ,
\end{equation}
which can be evaluated as described in \citep{Efstathiou1999} and gives
\begin{eqnarray}
r_{s}(a_{\rm{r}}) &=& \frac{4000}{\sqrt{\Omega_b h^2}} \frac{\sqrt{a_{eq}}}{\sqrt{1+\eta_{\nu}}} \\
&& \ln \left\{ \frac{[1+R(z_{\rm{r}})]^{1/2}+[R(z_{\rm{r}})+R_{eq}]^{1/2}}{1+\sqrt{R_{eq}}} \right\} \, \mathrm{Mpc}\nonumber
\end{eqnarray}
where $\eta_{\nu} = 0.6813$ is the ratio of the energy density in neutrinos to the energy
in photons. The parameter $R= 3\rho_{b}/4\rho_{\gamma}$ is numerically obtained
\begin{equation}
R(a)=30496 \Omega_b h^2 a.
\end{equation}
The scale factor at which radiation and matter have equal densities is
\begin{equation}
a_{eq}^{-1} = 24185 \left( \frac{1.6813}{1+\eta_{\nu}} \Omega_m h^2 \right),
\end{equation}
and the redshift of recombination $z_r$ is given by \citep{Hu1996} in the following
fitting formula
\begin{equation}
z_{\rm{r}} = 1048 \left(1+0.00124 (\Omega_{b} h^2)^{-0.738}\right)(1+g_{1} \, (\Omega_{m} h^2)^{g_{2}}) ,
\end{equation}
being
\begin{equation}
g_1 = 0.0783 (\Omega_{b} h^2)^{-0.238} \left( 1+ 39.5 \, (\Omega_{b} h^2)^{0.763} \right)^{-1},
\end{equation}
and
\begin{equation}
g_2 = 0.560 \left( 1+ 21.1 (\Omega_{b} h^2)^{1.81} \, \right)^{-1}.
\end{equation}


The $\chi^2$ function for BAO mock data is now defined in Eq. \ref{gegant}
\begin{figure*}
\normalsize
\setcounter{MYtempeqncnt}{\value{equation}}
\setcounter{equation}{44}
\begin{eqnarray}
\chi^{2}_{\mathrm{BAO}}(\boldsymbol{\theta}) &=& \frac{1}{1-\rho_{y,y'}^2}\left[\sum^{N_{mock}}_{j =
1} \frac{(y(z_{j}, \Omega_m, \Omega_b; \boldsymbol{\theta}) -
y_{mock}(z_{j}))^{2}}{\sigma^{2}_{y,j}}+\sum^{N_{mock}}_{j =
1} \frac{(y'(z_{j}, \Omega_m, \Omega_b; \boldsymbol{\theta}) -
y'_{mock}(z_{j}))^{2}}{\sigma^{2}_{y',j}}\right.  \nonumber \\
&-&2\left.\rho_{y,y'} \sum^{N_{mock}}_{j =
1} \frac{
(y(z_{j}, \Omega_m, \Omega_b; \boldsymbol{\theta}) -
y_{mock}(z_{j}))
(y'(z_{j}, \Omega_m, \Omega_b; \boldsymbol{\theta}) -
y'_{mock}(z_{j}))}{\sigma_{y,j}\sigma_{y',j}}\right]\nonumber\\
 \, 
\label{gegant}\end{eqnarray}
\setcounter{equation}{\value{MYtempeqncnt}}
\hrulefill
\vspace*{4pt}
\end{figure*}
\addtocounter{equation}{1}, where $N_{mode}$ is the number of mock data, in this case $4$. And $\boldsymbol{\theta}={\theta_1,\theta_2,...}$ are the dark energy model parameters. Here we have conveniently accounted for the slight degree of correlation existing between $y$ and $y'$,
and as suggested in \citep{Seo:2007ns}
we will fix for our calculations $\rho_{y,y'}=0.4$.



\begin{table*}
\centering
\caption{Mock BAO data from }\label{tab:JPASdata}
\begin{tabular}{c@{~~~}c@{~~~}c}
\hline\hline\\[-0.3cm]
$\mathbf{z}$ & $\mathbf{y}$& $\mathbf{y'}$ \\ \\[-0.3cm] \hline \\
$0.6$ & $15.005\pm0.117$ & $21.401\pm0.291$\\
$0.8$ & $19.221\pm0.109$ & $18.953\pm0.186$\\
$1.0$ & $22.708\pm0.102$ & $17.012\pm0.137$\\
$1.2$ & $25.762\pm0.098$ & $15.098\pm0.101$\\
$1.4$ & $28.635\pm0.096$ & $13.421\pm0.079$\\
$1.6$ & $31.126\pm0.101$ & $12.213\pm0.069$\\
$1.8$ & $33.540\pm0.107$ & $10.987\pm0.061$\\
$2.0$ & $35.357\pm0.112$ & $10.094\pm0.055$\\ \\[-0.23cm] 
\hline \hline
\end{tabular}
\end{table*}

\subsection{Type Ia Supernovae}
\label{sec:SNdata}
Type Ia supernovae (SN) are the explosions that take place at late stages of the stellar evolution. They have been recognized as a powerful probe of cosmological dynamics, as they give a good  measure of the cosmological expansion rate.  Supernovae provided the first probe for the cosmological expansion \citep{Riess1998,Perlmutter1999} and are considered as standard candles \citep{Leibundgut2001}.

The statistical analysis of  such SN samples rests on the definition of the modulus distance:
\begin{equation}
\mu(z_{j}) = 5 \log_{10} [ d_{L}(z_{j}, \boldsymbol{\theta}) ]+\mu_0,
\end{equation}
where $d_{L}(z_{j},  \boldsymbol{\theta})$ is the luminosity distance:
\begin{equation}\label{eq:dl_H}
d_{L}(z, \boldsymbol{\theta}) = (1+z) \ \int_{0}^{z} \mathrm{d}z'
\frac{1}{H(z', \boldsymbol{\theta})}.
\end{equation}
The best fits for the parameters of a given model can be obtained by minimizing
\be
\chi^2_{\rm SN}(\mu_0,\btheta)=\sum_{j=1}\frac{(\mu_{\rm th}(z_j;\mu_0,\btheta)-\mu_{\rm obs}(z_j))^2}{\sigma_{\mu,j}^2},
\ee
where $\sigma_{\mu,j}$ are the measurement variances.  Here we have neglected correlations between measurements at different redshifts,
as they are typically small; as it is well known, doing this just induces a slight degradation in the constraints, and therefore it is simply a conservative and acceptable procedure. 

In our  $\chi^2_{\rm SN}$ we have a
nuisance parameter, $\mu_0$, which encodes the Hubble parameter and
the absolute magnitude $M$, and has to be marginalized over. However,
when one works with an homogeneous data sample \citep{Nesseris2004}, an
alternative method is used for that purpose. This method consists in maximizing the likelihood by minimizing $\chi^2$ with respect to $\mu_0$ \citep{Elgaroy2006}.
Then one can rewrite the $\chi^2_{\rm SN}(\mu_0,\btheta)$ as
\begin{equation}
\chi^2_{\rm SN}(\btheta)=c_1-2c_2\mu_0+c_3\mu_0^2
\end{equation}
being
\begin{eqnarray}
&&c_1=\sum_{j=1}\frac{\left(\mu_{\rm obs}(z_j)-5\log_{10}d_L\left(z_j;\btheta\right)\right)^2}{\sigma_{\mu,j}^2}\quad\\
&&c_2=\sum_{j=1}\frac{\mu_{\rm obs}(z_j)-5\log_{10}d_L\left(z_j;\btheta\right)}{\sigma_{\mu,j}^2}\quad\\
&&c_3=\sum_{j=1}\frac{1}{\sigma_{\mu,j}^2}\quad.
\end{eqnarray}
With the minimization over $\mu_0$ one obtains an expression in terms
 of $c_2$ and $c_3$:
 \begin{equation}
 \mu_0=c_2/c_3,\label{eq:1}
\end{equation}
so the $\chi^2$ function takes the form
\begin{equation}
\tilde\chi^2_{\rm SN}(\btheta)=c_1-\frac{c_2^2}{c_3}. 
\end{equation}

\subsubsection{Union2}
Union2 \citep{Amanullah2010} is one of the largest type Ia Supernovae samples up
to date, and it consists of 557 SNIa and covers a redshift range from $0$ to
$1.4$. This sample has increased the number of well-measured Type Ia
supernovae of the previous Union \citep{Kowalski2008} at high
redshifts by the combination of different data sets.  The Union Sample
has been extended with six type Ia supernovae presented in
\citep{Amanullah2010},  the SNe from
\citep{Amanullah2008}, the low-z and intermediate-z data from
\citep{Hicken2009} and \citep{Holtzman2008} respectively.

This sample has been obtained after some improvements in the Union analysis chain: all light
curves have been fitted using a single light curve fitter (SALT2) to
eliminate differences and systematic errors have been directly
computed using the effect they have on the distance modulus.

\subsubsection{Mock data for a pre-WFIRST stage}
To create SN mock samples we have reproduced the pre-WFIRST observational simulation as reported in 
\citep{Albrecht2009}, which has two population peaks, one at $z<0.1$ and the other at $0.6<z<0.7$; along with a very scarce population at $z>1.6$. Specifically,
the redshift distribution suggested in \citep{Albrecht2009} is reproduced  here in
Table~\ref{tab:JDEM}.

\begin{table*}
\begin{center}
\caption{Redshift distribution of pre-WFIRST SN samples}\label{tab:JDEM}
\begin{tabular}{cc}
\hline
\hline
{redshift bin}  &  {$\#$ of SN}  \\
\hline
\hline
$<0.1$     & $500$  \\
$0.1-0.2$  & $200$  \\
$0.2-0.3$  & $320$  \\
$0.3-0.4$  & $445$  \\
$0.4-0.5$  & $580$  \\
$0.5-0.6$  & $660$  \\
$0.6-0.7$  & $700$  \\
$0.7-0.8$  & $670$  \\
$0.8-0.9$  & $110$  \\
$0.9-1.0$  & $80$   \\
$1.0-1.1$  & $25$   \\
$1.1-1.2$  & $16$   \\
$1.2-1.3$  & $16$   \\
$1.3-1.4$  & $4$    \\
$1.4-1.5$  & $4$    \\
$1.5-1.6$  & $4$    \\
$>1.6$     & $4$    \\
\hline
\hline
\end{tabular}
\end{center}
\end{table*}

The formulae for errors on SN magnitudes that we use follows from a prescription used in the binning approach \citep{Kim2004,Upadhye:2004hh,
Ishak:2005xp},
in which they are calculated as follows:
\begin{equation}\label{eq:error}
\sigma_{m}^{\mathit{eff}} = \sqrt{\sigma_{int}^2 + \sigma_{pec}^2 + \sigma_{syst}^2} \; ,
\end{equation}
where:
\begin{itemize}
 \item $\sigma_{int} = 0.15$ is the intrinsic dispersion in magnitude per SN,
       assumed to be constant and independent of redshift for all  well-designed surveys;
 \item $\sigma_{pec} = {5 \sigma_{v}}/{(\ln(10) c z)}$ is the error due to the uncertainty
       in the SN peculiar velocity, with $\sigma_{v} = 500$ km$/$s, $c$ is the velocity of light  and $z$ is the redshift for any SN;
 \item $\sigma_{syst} = 0.02 ({z}/{z_{max}})$ is the floor uncertainty related to all the irreducible
       systematic errors with cannot be reduced statistically by increasing the number of observations.
       The value $0.02$ is conservative from the perspective of what space-based missions could achieve. Those
       are precisely the resources expected to provide high-redshift SN, which are in turn the ones
       in which the systematic error term is expected to contribute. Note as well that $z_{max}$ is the maximum 
       observable redshift in
       the considered mission and this linear term in redshift is used to account for
       the dependence with redshift of many of the possible systematic error sources (for example the
       Malmquist bias or gravitational lensing effects).
\end{itemize}
We have included some extra (though very slight) noise, and then we have checked that our mock data
are compliant with the main features of the Union2 sample.
The fiducial model used is  again the wcdm+sz+lens case from WMAP7-year
as quoted above , which has $\Omega_m= 0.26\pm0.099$ and is phantom-like with $w= -1.12\pm0.43$.
The $\chi^2$ function for the SN mock data has been constructed as described in  Sec.~(\ref{sec:SNdata}). 
Then we have let 
\textit{iCosmo} generate the $d_L$ values for the redshifts in the table a large number of universes drawn randomly and normally distributed around the fiducial one, and finally we have performed a reduction to give our mock sample.

\begin{figure*}
\begin{center}
\includegraphics[width=0.5\textwidth]{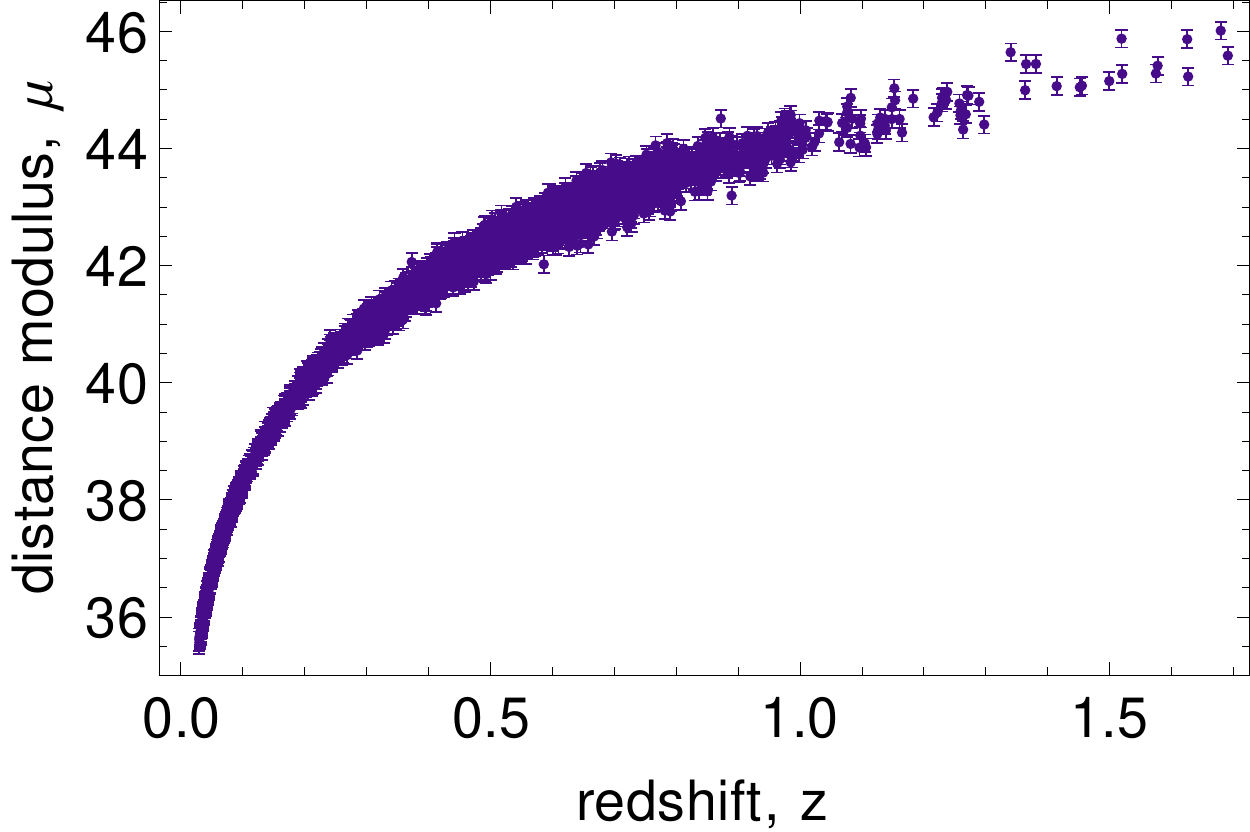}
\caption{\label{fig:snemock} 
Mock data obtained with iCosmo for the pre-WFIRST specifications.}
 \end{center}\end{figure*}

\section{Results}
Following Bayesian Statistics, we have inferred the values $\Omega_m$ and the dark energy parameters for the models considered. For further details   see \ref{Par_est}. 
We report our findings in two main ways: on the one hand we present our best fits, errors
and derived quantities in 
Tables \ref{Uniontable}, \ref{newdatatable};  on the other hand we present credible contours obtained after a numerical marginalization over $\Omega_m$, where we have approximate the likelihood as a Gaussian and then analytically marginalize, see \citep{Taylor2010} and references therein.
As mentioned, we have considered the combination of real and mock SN and BAO data as discussed above, and in addition 
we have introduced in all cases a Gaussian prior on $\Omega_m$ and $\Omega_b$ deduced 
in \citep{Burigana2010} as a forecast analysis of Planck: $\Omega_mh^2=0.1308\pm0.0008$, $\Omega_b=0.0223$ with $h=0.742$ as is given by \citep{Riess2009}. 

The presence of the prior leads to uniformity in the best fit value of
$\Omega_m$  thus minorating its influence in the dark energy constraints. This trick, together with the 
marginalization, allow to focus the discussion on the dark energy parameters thus offering a clearer picture of
what each parametrization can offer.

\begin{table*}
\caption{Constraints on dark energy parameters and derived quantities from current data.\label{Uniontable}}
\begin{center}
\begin{tabular}{@{~~}c@{~~}c@{~~}c@{~~}c@{~~}c@{~~}c@{~~}c@{~~}c@{~~}}
\hline\hline\\[-0.2cm]
Model &  $\chi^2_{min}$   &
$\Omega_m$ & Dark energy parameters&  ${\rm FoM_{Wang}}$ & $\rho_{1,2}$ & DIC \\
&&&&$({\rm FoM_{DETF}})$  \\\\[-0.2cm] \hline \\
CPL & $544.91$  & $0.315\pm0.033$ & $w_0=-1.033\pm0.180$ &  $14.16$ & $-0.930$ & $10.30$ \\ 

& & & $w_a=-0.742\pm1.520$ & $(14.16)$         &           \\ \\
Wang & $544.91$  & $0.315\pm0.033$ & $w_0=-1.033\pm0.180$ &  $42.47$ &  $-0.792$ & $8.11$ \\ 

& & & $w_{0.5}=-1.281\pm0.361$ &   $(14.16)$   &           \\ \\
Chebychev Polynomial & $544.96$ & $0.314\pm 0.032$ & $w_0=-1.049\pm 0.163$ & $43.63$ & $-0.785$ & $8.02$ \\ 

&  &  & $w_{0.5}=-1.268\pm 0.373$   & $(11.67)$   &           \\ \\
Conventional Polynomial & $544.98$  & $0.314\pm0.032$ & $w_0=-1.055\pm0.157$ & $44.11$ & $-0.780$ & $7.98$ \\ 

&  &   & $w_{0.5}=-1.263\pm0.377$  &  $(9.80)$  &           \\ \\
\hline \hline
\end{tabular}
\end{center}
\end{table*}

\begin{table*}
\caption{Constraints on dark energy parameters and derived quantities from simulated data. \label{newdatatable}}
\begin{center}
\begin{tabular}{@{~~}c@{~~}c@{~~}c@{~~}c@{~~}c@{~~}c@{~~}c@{~~}c@{~~}}
\hline\hline\\[-0.2cm]
Model &  $\chi^2_{min}$   &
$\Omega_m$ & Dark energy parameters & ${\rm FoM_{Wang}}$ & $\rho_{1,2}$ & DIC \\
&&&&$({\rm FoM_{DETF}})$  \\\\[-0.2cm] \hline \\
CPL & $5320.38$  & $0.269\pm0.005$ & $w_0=-1.151\pm0.041$ & $616.25$ & $-0.946$ & $21.44$ \\
&             &                   & $w_a=0.244\pm0.207$ &  ($616.25$)      &           \\ \\
Wang & $5320.38$  & $0.269\pm0.005$ & $w_0=-1.151\pm0.041$ & $1848.70$ &  $-0.712$ & $11.69$ \\
&             &                   & $w_{0.5}=-1.069\pm0.031$ &  ($616.25$)       &           \\ \\
Chebychev polynomial  & $5320.43$  & $0.269\pm0.005$ & $w_0=-1.140\pm0.035$ & $2065.28$ & $-0.666$ & $11.38$ \\ 
&             &                   & $w_{0.5}=-1.073\pm0.029$  & ($504.85$)   &         \\  \\ 
Conventional polynomial & $5320.45$ & $0.269\pm0.005$ & $w_0=-1.137\pm0.033$ & $2147.57$ & $-0.648$ & $11.29$ \\ 
&            &                   & $w_{0.5}=-1.075\pm0.028$  & ($477.24$)   &           \\ \\
\hline \hline
\end{tabular}
\end{center}
\end{table*}

Let us now examine our results using  different criteria.
First of all uncertainties in percentage terms on the dark energy values can be considered. 
The conventional polynomial turns out to give the lowest percentage error on $w_0$ for
both current and recent data, and the second best is the Chebyshev setting, although we only get a marginal difference with
respect to the conventional polynomial and Wang's model.  The situation gets reversed, though, for the percentage error on $w_{0.5}$ , and Wang's parametrization is the best performer, but again differences
are small when compared to our parametrizations (CPL gets excluded from this discussion
item as the parameter $w_{0.5}$ is not considered).

Note  that  the substantial  reduction on percentual errors when one moves from current to mock data, typically they become almost three times smaller. Up to some degree this may be a consequence of the use of a fiducial model in the simulated
data, probably real future data will be significantly better than current ones but perhaps
not as remarkably as our synthetic data. 

Summarizing, this section of the analysis,  we see that in general 
choosing $w_0$ and $w_{0.5}$ as the parameters to constrain is worthy as percentual errors are low. 
In our two new parametrizations are indeed valuable from this particular perspective, but are they 
as valuable when one considers other criteria?

Another interesting point of view to interpret our results is 
Pearson's correlation coefficient (for the dark energy), defined as
\be
\rho_{1,2}=\frac{\sigma_{12}^2}{\sigma_1\sigma_2}
\ee
which can be used to study the lineal correlation between either two dark
energy parameters. Here $\sigma_{12}$ stands generically for the non-diagonal element of the covariance 
matrix for the parameters $1$ and $2$. A value of $\rho_{1,2}$ close to zero will tell us 
there is no linear correlation between them, and lowering correlation typically improves 
constraints, so  if a given parametrization achieves that goal naturally then 
it will  most likely provide an overall worthy scenario. 

A related magnitude is the (frequentist) Figure-of-Merit (FoM), which has been defined in two slightly different but related ways in the literature. We present both and discuss their specific features and report their values for
the benefit of the readers keener to one version or the other.
Following the corresponding references we use subscripts to differentiate them.
According to \citep{Wang2008a}
\be
{\rm FoM_{Wang}}=\frac{1}{\sqrt{\det {\mathbf C}(c_1,c_2,c_3,...)}},
\label{eq:FoM}
\ee
where $ {\mathbf C}(c_1,c_2,c_3,...)$ is the covariance matrix of the corresponding ${c_i}$ dark energy parameters one is concerned
with.
As explained in \citep{Wang2008a}, this specific definition has two advantages: 
firstly it is easy to calculate for either real or simulated data, and secondly it 
has an easy physical interpretation. The FoM penalizes experiments that yield 
very correlated estimates for the dark energy parameters. Hence the FoM is
larger when the dark energy parameters  ${c_i}$ are chosen such that they are minimally correlated with each other.
But obviously, if the model considered does not have a low degree of correlation {\it per se} one will be giving a poor
estimation of the performance of the observational tests. 

A closely related, and perhaps more popular version of the FoM was proposed by the ``Dark Energy Task Force'' (see 
\citep{Albrecht2009} for a recent revision of the topic), and 
it simply reads
\be
{\rm FoM_{DETF}}=\frac{1}{\sqrt{\det {\mathbf C}(w_0,w_a)}}.
\label{eq:FoMdetf}
\ee
Thus, the main difference with respect to the ${\rm FoM_{Wang}}$ is that in this view the two relevant parameters are $w_0$ and $w_a$,
and not any others. However, $w_a$ is a parameter informing about early asymptotics, which is a region with extremely
hard observational access and physical interpretation, and thus, the debate persists of whether
a FoM using that parameter may not be given artificially large results.

According to Appendix \ref{ap:C}, the two FoMs are linearly related:
\be{\rm FoM}_{\rm DETF}=m_3 {\rm FoM}_{\rm {Wang}}
\ee
with $m_3= 1/3, 2/9, 11/45$ for Wang, Chebychev polynomial, Conventional polynomial models respectively.

Our results on the FoM get summarized very simply:
the conventional polynomial  has the largest value of the ${\rm FoM_{Wang}}$, whereas
the results get complete reversed for the ${\rm FoM_{DETF}}$. In addition, the ratios between the different FoMs are very similar
for real and synthetic data. However, as follows from our discussion before,
the price paid to get a larger FoM (by a redefinition) it to waive the importance of correlation, and further
investigations and reflections would be needed to provide a definite solution to this debate, which is, on the other hand,
out of the scope of this paper.

\begin{figure*}

\includegraphics[width=0.81\textwidth]{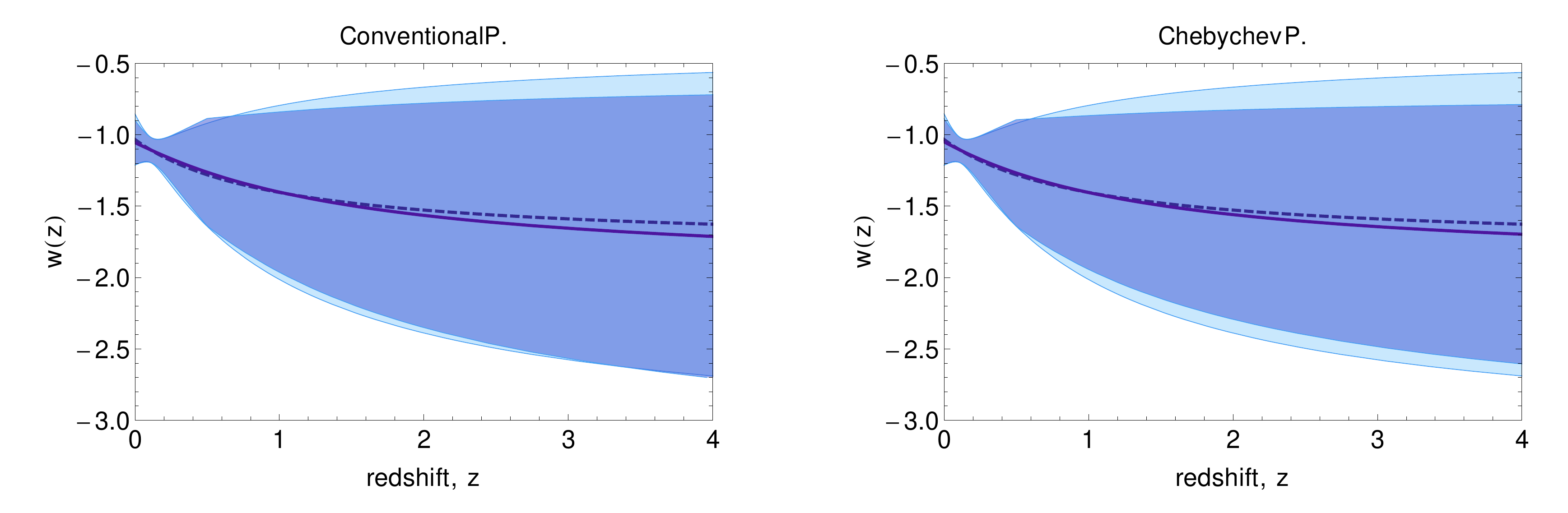}
\label{fig:wr}

\includegraphics[width=0.81\textwidth]{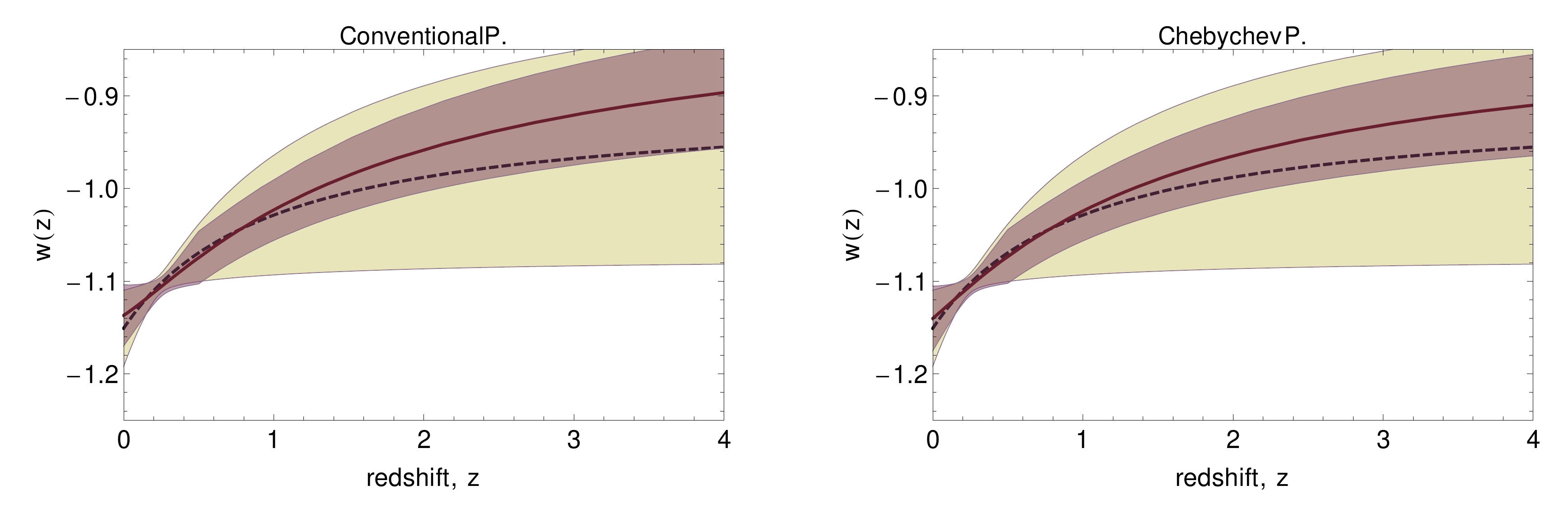}
\label{fig:wm}

\caption{\label{fig:wtot} Total $w$  and $1\sigma$ error bands for our best fits for real (a) and mock data (b) compared with the results for
the CPL parametrization (dashed lines and light contours).}
\end{figure*}

For both sorts of  data the conventional polynomial parametrization is naturally less correlated than all three others. 
The second best is the Chebyshev one, and all three are considerably less correlated than CPL.
Nevertheless, even though Wang's second parameter $w_c\equiv w_{0.5}$ is chosen for
the low degree of correlation with $w_0$, a better
choice for that purpose would be the dark energy EOS parameter evaluated at a lower redshift (say $z\sim 0.25$). This was already
pointed out in \citep{Wang2008a} and confirmed in \citep{EscamillaRivera:2011qb}. 
In agreement with our discussion the degree of correlation at a certain low redshift in our two new parametrizations
immediately drives to very narrow errors on the total $w$ at that location, as reflected in Fig. \ref{fig:wtot},  where
we represent the total $w$ for our best fits and the $1\sigma$ error bands. In fact these two figures
serve the additional purpose of illustrating the sort of evolution described our parametrizations.  
Nevertheless, as correlation is a topic worth of further consideration we elaborate
in Appendix \ref{ap:B}.

Finally, the last criterion we resort to is the Bayesian deviance information criterion (DIC). This is a very interesting way to
examine results in this context, in particular when one finds marginal differences in the $\chi^2$ values. This criterion accounts
for the dependence of $\chi^2$ on our parameters, but it also somehow involves the correlation and the percentual errors, thus
turning out to be far more informative and having more discerning power. When applied for model selection, as we put forward,
the setting with the lowest DIC is in principle the best. 
In our case, as we are mainly concerned with dark energy
issues given that $\Omega_m$ is very tightly constrained we construct our DIC starting from a $\chi^2$ marginalized over $\Omega_m$.
Namely 
\be
{\rm DIC}=2\overline {\widehat{\chi^2}(\btheta)}-{\widehat{\chi^2}}(\btheta_{bf}),
\ee
where 
\be\widehat{\chi^2}=-2\log\left(\int_0^1\exp(-\chi^2/2)d\Omega_m\right).\ee

The behaviour of the DIC follows the same pattern as the FoM (although the FoM is not so convenient as a model selection criterion
as it does not involve $\chi^2$). Basically, the conventional polynomial model is the best one, then we have the Chebyshev polynomial
model, then Wang's scenario, and finally, the CPL model closes the ranking with the highest DIC by far.

\begin{figure*}
\begin{center}
\includegraphics[width=0.70\textwidth]{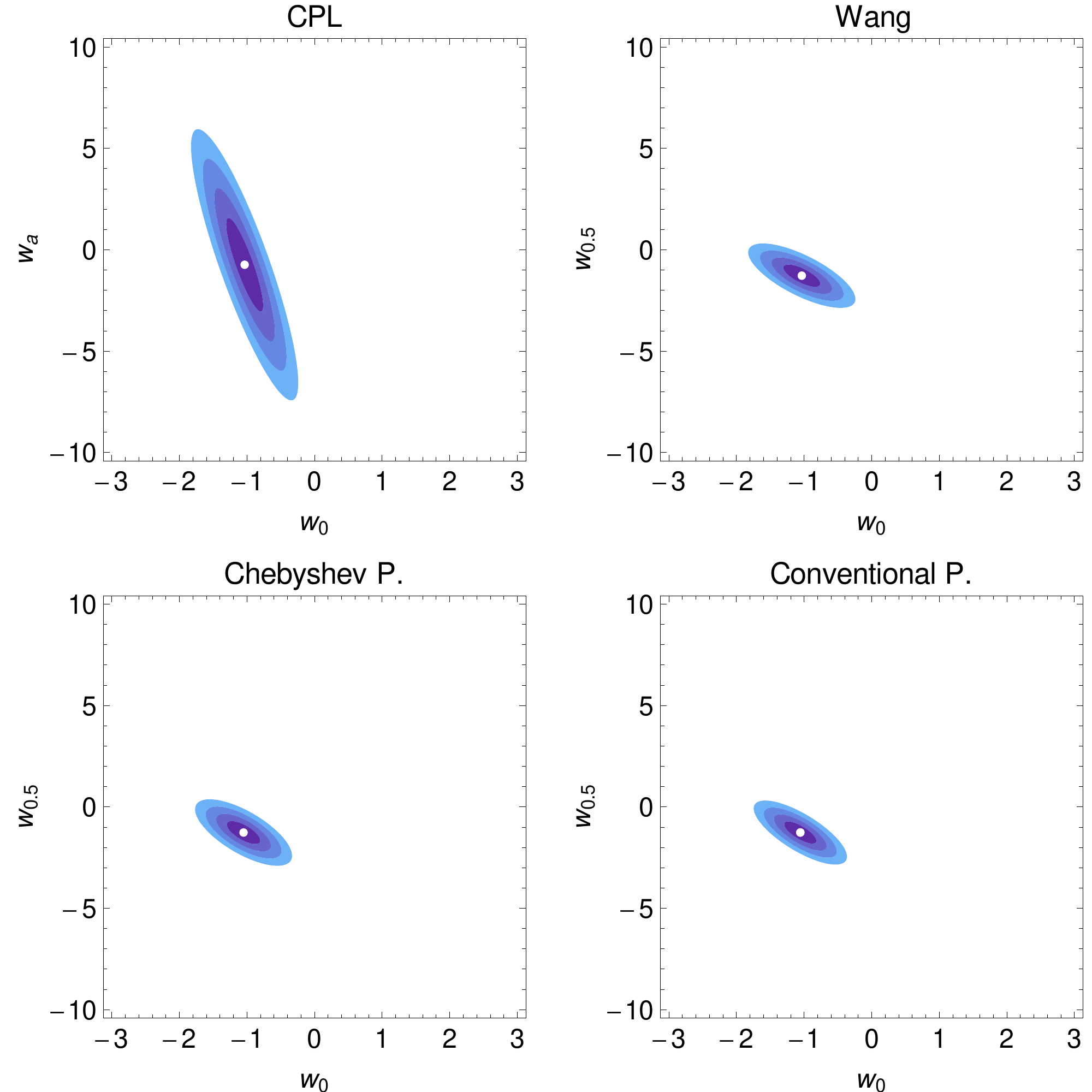}
\caption{\label{fig:gridonakUNION} Confidence contours  for the four parametrizations using 
current data.}
 \end{center}\end{figure*}
 
\begin{figure*}
\begin{center}
\includegraphics[width=0.70\textwidth]{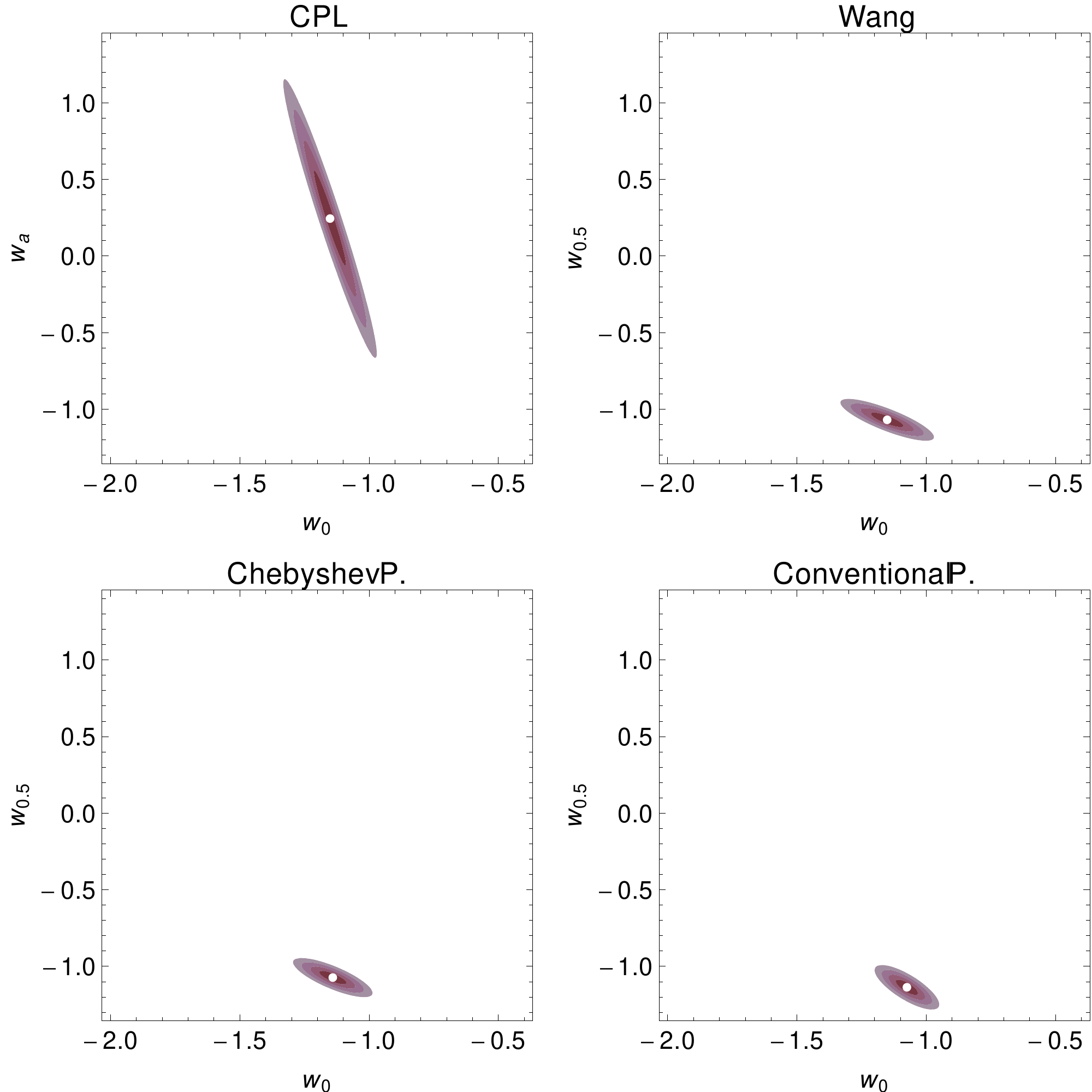}

\caption{\label{fig:gridonakMOCK}  Confidence contours  for the four parametrizations   using 
simulated data.}
 \end{center}\end{figure*}

Finally, we can see that the FoM values obtained with the mock data are typically better than those for currently available data. This fact proves the capability of the forthcoming surveys to describe the evolutionary features of dark energy.

\section{Conclusions}
Parametrizing dark energy in a phenomenological way offers the possibility of progressing in the characterization
of this main component of the Universe even though its origin has not been yet unveiled from a theoretical perspective.
The extensive prior knowledge available on the topic hints that compliance with relevant astrophysical 
data favours parametrizations which are smooth and have two parameters only. A second order requirement is
restraint at high redshifts in the sense that dark energy should never redshift slower than matter, but neither
display an significant blueshift.

Given those guidelines, in this work we present two new polynomial parametrizations of dark energy and we
pay attention to some particular aspects which are important for model selection. On the one
hand we examine the degree of correlation between the two parameters chosen in our proposal; 
specifically those are the EOS values at $z=0$ and $z=0.5$, which have been already shown to provide
a convenient way to revisit  and have already been shown to provide a preferred way to reconsider.
On the other hand we resort to a genuinely Bayesian model selection criterion, the deviance information 
criterion (DIC) for a better account of the improvements that our parametrizations represent. 

The astronomical tests we perform make use of SN luminosity data and the typical BAO related scales, 
so we are focusing on low redshift probes, although our parametrizations are suitable (as they
are well behaved at high redshifts) for the analysis of CMB constraints which we hope to address in the future. 

The conclusions that emerge from our analysis is that our new parametrizations perform better in the sense
that they allow to obtain tighter constraints on the dark energy EOS at present and its derivative, and
they are also favoured by the  statistical indicator we mention above (DIC). The main reason why we feel they fare better
than their competitors is that they represent rather flexible perturbations of the $\Lambda$CDM scenario, which in many respects still remains
the best description of the accelerated universe, and not surprisingly is often referred to as the concordance model. 
In contrast, either CPL or Wang are models which when taken as perturbations of $\Lambda$CDM are left with one free parameter only, and
then one could expect less ability to accommodate themselves to the data. 

Note as well that the dataset we consider are not only some currently available, but we also simulate surveys to come,
in particular, it is of interest our provision of synthetic measurements of the radial and transversal BAO scales as expected to be obtained with the EUCLID spectroscopic survey, which as expected help us conclude that future surveys will decrease considerably
our degree of ignorance about dark energy evolution by providing far tighter constraints that may eventually lead to the conclusion
that a cosmological constant is not the preferred candidate for the dark energy.

\section*{Acknowledgements} 
Throughout this work we have benefited from illuminating conversations with Vincenzo Salzano and Narciso Ben\'\i tez, and
we also acknowledge fruitful exchanges with Adam Amara, Bruce Bassett, Karl Glazebrook, Christopher Gordon, Thomas Kitching,  David Parkinson,
Anais Rassat, and Yun Wang.
Irene Sendra holds a PhD FPI fellowship-contract from the Spanish Ministry of Science and Innovation and
Ruth Lazkoz  and Irene Sendra are supported by the Spanish Ministry of Science and Innovation through research projects FIS2010-15492, Consolider EPI CSD2010-00064
and also by the Basque Government through the special research action KAFEA during the initial stage of this work.

\appendix
\section{Parameter estimation}\label{Par_est}
\label{ap:A}

The likelihood function, ${\cal L}(\textbf{d}\vert \btheta,{\cal M})$, is defined up to proportionality, as the probability of measuring the data  $\textbf{d}=\left\{d_1,\dots, d_n\right\}$ given the model ${\cal M}$ and its parameters taking  the values $\btheta=\left\{\theta_1,\dots,\theta_{\nu}\right\}$ \citep{jussi,Denison2002}. 

Despite our aim to keep the discussion in this section as general as possible, when we analyse particular datasets we will assume, as usual, that the measurements are normally distributed around their true value, so that
\begin{equation}\label{gauss lik}
{\cal L}(\textbf{d}\vert \btheta,{\cal M})\propto e^{-\chi^2(\btheta)/2}.
\end{equation}

The probability density function $p(\btheta \vert \textbf{d}, {\cal M})$  of the parameters to have values
$\btheta$ for the data, $\textbf{d}$, under the assumption that the true  model is ${\cal M}$ is provided by Bayes' theorem \citep{jussi}

\begin{equation}
p(\btheta \vert \textbf{d},{\cal M})= \frac{{\cal L}(\textbf{d}\vert \btheta,{\cal M})\pi(\btheta,{\cal M})}{\int{\cal L}(\textbf{d}\vert \btheta,{\cal M})\pi(\btheta,{\cal M})d\btheta},
\end{equation}
where $p(\btheta \vert \textbf{d},{\cal M})$ and $\pi(\btheta,{\cal M})$ are the posterior and prior probability density functions (pdf) respectively \citep{jussi,Trotta2004,Trotta2007,cousins,Denison2002}. The prior pdf encodes all previous knowledge about the parameters before the observational data have been collected. It can be regarded as a subjective procedure, but its use is compulsory in  the Bayesian framework, which is the approach used in theoretical frameworks
where only one particular realization of the measurement is available.

Parameter estimation in the Bayesian framework is based on maximizing the posterior pdf $p(\btheta \vert \textbf{d},{\cal M})$, 
whereas in a ``strict''   frequentist approach  one just maximizes $ {\cal L}(\textbf{d} \vert \btheta,{\cal M})$. 
When one uses flat priors in the Bayesian approach then the same conclusions are drawn from both approaches and then the difference turns 
to be conceptual only \citep{Trotta2004,Trotta2007,trotta3}. If the measured observables are independent form each other and Gaussian 
distributed around their true value, $\textbf{d}(\btheta)$, with a covariance matrix ${\mathbf C}$, given by the experimental errors, 
maximizing ${\cal L}$ is equivalent to minimizing the chi-square function

\be
\chi^2(\btheta)\equiv\left(\textbf{d}^{obs}-\textbf{d}(\btheta)\right){\mathbf C}^{-1}\left(\textbf{d}^{obs}-\textbf{d}(\btheta)\right)^{T}
\ee

and for uncorrelated data $C_{ij}=\delta_{ij}\sigma_i^2$,
\be
\chi^2(\btheta)\equiv\sum_{i=1}^n\left(\frac{d^{obs}_i-d_i(\btheta)}{\sigma^{obs}_i}\right)^2.
\ee

The second step toward constraining parameters satisfactorily is to construct credible intervals \citep{Trotta2004} which measure the degree of confidence that a certain data was generated by parameters belonging to the estimated interval.
which gives the best fit parameters
In the Bayesian approach, the credible intervals are drawn around the maximum likelihood point, $\btheta_{bf}$. After obtaining it by the minimization of the $\chi^2(\btheta)$, the boundaries of the region containing $100n\%$ of likelihood are determined as the values of the parameters for which $\chi^2$ has increased by a certain quantity
\be
\chi^2-\chi^2_{min}=\Delta_{\nu,n}
\ee

with

\be
n=1-\frac{\displaystyle{\int^\infty_{\Delta_{\frac{\nu,n}{2}}}t^{\frac{\nu}{2}-1}e^{-t}dt}}{\displaystyle{\int^\infty_{0}t^{\frac{\nu}{2}-1}e{-t}dt}}=1-\frac{\displaystyle{\Gamma\left(\frac{\nu}{2},\frac{\Delta_{\nu,k}}{2}\right)}}{\displaystyle{\Gamma\left(\frac{\nu}{2}\right)}}
\ee
where $\Gamma\left({\nu}/{2},{\Delta_{\nu,k}}/{2}\right)$ is the incomplete $\Gamma$ function \citep{Lazkoz2005}, \citep{Press1992}.

The $1\sigma$ and $2\sigma$ errors of the parameter $\theta_i$ are given by the $68.30\%$ and $95.45\%$ credible interval contours, respectively. The upper limit is the maximum value of the contour and the lower one the minimum one.

\section{Pivot computation}
\label{ap:B}
\subsection{Wang's parametrization}
The change of variables transforming the CPL parametrization into Wang's is
\be
w_c=w_0+(1-a_c)w_a,\label{wc}
\ee
where $w_c$ is the dark energy parameter EOS at the pivot redshift $z_c$ for which some correlation related (good) feature is achieved.
Wang's choice $w_c=0.5$ leads to a low correlation situation, whereas we are letting $w_c$ stand for the value associated with the
lowest correlation possible, so our scope is more general.

Using error propagation one gets
\be
\sigma_c^2=\sigma_0^2+(1-a_c)\sigma_a^2+2(1-a_c) \sigma_{12},
\ee
directly from Eq. \ref{wc}. In the remainder we will drop the absolute value given that 
we are only interested in a $a_c<1$ situation.
In contrast, if one solves Eq. \ref{wc} for $w_a$ and then applies on it error propagation the result
\be
\sigma_a^2=\frac{\sigma_0^2+\sigma_c^2-2 \sigma_{0c}}{(1-a_c)^2}
\ee
follows. Combining our results above
\be\sigma_0^2+(1-a_c)\sigma_{12}=\sigma_{0c}\label{correl}
\ee
is obtained, and then one deduces that total decorrelation ($\sigma_{0c}=0$)
is achieved for
\be
a_c=1+\sigma_0^2/\sigma_{12},
\ee
which corresponds for our results to $z_c=0.16,0.26$ respectively for real and synthetic data.
Note that according to our notation
$\sigma_{12}$ denotes in this case the non-diagonal element of the covariance matrix of $w_0$ and $w_a$,
whereas 
$\sigma_{0c}$ denotes the non-diagonal element of the covariance matrix between $w_0$ and $w_c$.

But if one is just demanding a situation where $\sigma_{0c}<\sigma_{12}$ it is very easy
to deduce from Eq. \ref{correl} that such condition is met whenever
\be
a_c>\sigma_0^2/\sigma_{12}.
\ee

\subsection{Chebyshev polynomial}
Provided $w_{0.5}=w\vert_{z=0.5}$, let us adopt for $w_c$  the most general definition along the ideas discussed in their
previous subsection. For our Chebyshev polynomial parametrization we have
\bea
w_c=&&\frac{1}{11} \left(-6 a_c^2 (4w_0-3
  w_{0.5}+1)+a_c (73w_0-63w_{0.5}+\right.\nonumber\\
  &&\left.10)-38
  w_0+45w_{0.5}-4\right).
   \eea
   Following the same straightforward calculation as for the previous case 
   we conclude there will be minimal correlation for
  \be a_c=\frac{\sqrt{729\sigma_{12}^2-2142\sigma_{12}
   \sigma_0^2+1681 \sigma_0^4}+63\sigma_{12}-73
   \sigma_0^2}{12 (3\sigma_{12}-4 \sigma_0^2)},
   \ee
      which corresponds to $z_c=0.17,0.28$ respectively for real and synthetic data.
 Sticking to our notation  here
$\sigma_{12}$ denotes the non-diagonal element of the covariance matrix of $w_0$ and $w_{0.5}$, and
$\sigma_{0c}$ has the same meaning as in the case before.

   But again one can just want to meet the less restrictive requirement  $\sigma_{0c}<\sigma_{12}$, 
   which follows provided 
   \be
(3 a_c-2) (6 a_c-17)\sigma_{12}<
\left(24 a_c^2-73 a_c+38\right) \sigma_0^2\,.
   \ee
   ~
 \subsection{Conventional polynomial}
Finally, following the same steps as before, but this time considering the conventional
polynomial parametrization we have
\bea
w_c=&&\frac{1}{4} (4 (2-3 a_c) (a_c-2)w_0+9
   (a_c-2) (a_c-1)w_{0.5}+\nonumber\\&&(2-3 a_c)
   a_c+3 a_c-2).
\eea
The same route as for the two previous cases drives us to the conclusion that 
the minimal correlation situation is achieved for
  \be a_c=\frac{9 \sigma_{12}-8 \sigma_0^2}{9 \sigma_{12}-12
   \sigma_0^2},
   \ee
   which corresponds to $z_c=0.18,0.29$ respectively for real and synthetic data.
Here the meaning of   $\sigma_{12}$ and
$\sigma_{0c}$ is exactly the same of  the case before.
  
  Finally, one can deduce the condition for the less restrictive requirement  $\sigma_{0c}<\sigma_{12}$, 
   to happen is simply
   \be
      \frac{1}{4} (7-3 a_c) (2-3 a_c)\sigma_{12}<(2-3 a_c) (2-a_c)\sigma_0^2.
   \ee

\section{Equivalence between ${\rm FoM}_{\rm DETF}$ and ${\rm FoM}_{\rm{Wang}}$}
\label{ap:C}
 Following \citep{Coe2009} we get
 \begin{equation}
  \mathbf{C'^{-1}}={\mathbf M}^T\mathbf{C^{-1}}\mathbf{M}
 \end{equation}
where
   \begin{eqnarray}
   \mathbf{C}=\left(\begin{array}{cc}
\sigma_1^2  &     \sigma_{12}\\
 \sigma_{12} &  \sigma_2^2 \\
\end{array} \right) \; .
\end{eqnarray}
and  $M_{ij}=\partial p_i/\partial p'_j$. In our case
$\{p_1,p_2\}=\{w_0,w_{0.5}\}$ and $\{p'_1,p'_2\}=\{w_0,w_a\}$, and specifically
$w_{0.5}=m_1+m_2 w_0+m_3w_a$ with $m_1,m_2,m_3$ coefficients different in the different parametrizations considered other than
the CPL one, thus
   \begin{eqnarray}
   \mathbf{M}=\left(\begin{array}{cc}
1 &     0\\
m_2 &  m_3\\
\end{array} \right) \; .
\end{eqnarray}
It is a simple matter of algebra to see that
   \begin{eqnarray}\mathbf{C'}=\left(\begin{array}{c@{~~~}c}
\sigma_1^2  &     \displaystyle\frac{\sigma_{12}-m_2\sigma_1^2}{m_3}\\
\displaystyle \frac{\sigma_{12}-m_2\sigma_1^2}{m_3} &  \displaystyle\frac{m_2^2 \sigma_1^2 - 2 m_2 \sigma_{12} + \sigma_2^2}{m_3}\\
\end{array} \right) \; 
\end{eqnarray}
and $\sqrt{\det{C'}}=\sqrt{\det{C}}/m_3.$
Therefore we get the simple relationship between the two definitions of FoM
\begin{equation}{\rm FoM}_{\rm DETF}=m_3 {\rm FoM}_{\rm {Wang}}.
\end{equation}

\bibliographystyle{mn2e}
\bibliography{tesismore}

\bsp

\label{lastpage}
\end{document}